\documentclass[acus]{JAC2003}

\usepackage{graphicx,color}

\begin{document}
\title{{\normalsize \rm Particle Accelerator Conference (PAC\,07), Albuquerque, NM, 25-29 June 2007, THPMS082 \qquad NFMCC-doc-515} \\ MUON ACCELERATION TO 750 GeV IN THE TEVATRON TUNNEL FOR A  1.5 TeV
{\boldmath{$\,\mu^+ \mu^-$}} COLLIDER\thanks{Supported by DE-FG02-91ER40622 and 
DE-AC02-98CH10886.}}

\author{D.\,J. Summers\thanks{summers@phy.olemiss.edu}, L.\,M. Cremaldi, 
R. Godang\thanks{Also with University of South Alabama, Mobile, AL 36688, USA}, 
B.\,R. Kipapa, H.\,E. Rice, Univ. of Mississippi-Oxford, \\ University, MS 38677, USA;  \quad
R.\,B. Palmer, Brookhaven National Lab, Upton, NY 11973, USA}

\maketitle

\begin{abstract}
Muon acceleration from 30 to 750 GeV in 72 orbits using two rings in the 1000\,m radius Tevatron tunnel is explored. The first ring ramps at 400 Hz and accelerates muons from 30 to 400 GeV in 28 orbits using 14 GV of 1.3 GHz superconducting RF.  The ring duplicates the Fermilab 400 GeV  main ring FODO lattice, which had a 61\,m cell length. Muon survival is 80\%. The second ring accelerates muons from 400 to 750 GeV in 44 orbits using 8 GV of 1.3\,GHz superconducting RF.  The 30\,T/m main ring quadrupoles are lengthened 87\% to 3.3\,m. The four main ring dipoles in each half cell are replaced 
by three dipoles which ramp at 550 Hz from -1.8\,T to +1.8\,T
interleaved with two 8\,T fixed superconducting dipoles. The ramping and superconducting  dipoles oppose each other at 400 GeV and act in unison at 750 GeV. Muon survival is 92\%. Two mm copper wire, 0.28\,mm grain oriented silicon steel laminations, and a low duty cycle mitigate eddy current losses.  Low emittance muon bunches allow small aperatures and permit magnets to ramp with a few thousand volts. Little civil construction is required. The tunnel exists.
\end{abstract}

\section{MUON COLLIDER INTRODUCTION}

A muon collider\,\cite{Raja} can do s-channel scans to try to split
the $H^0\!/\!A^0$ Higgs doublet\,\cite{Barger}.
At a 1.5 TeV frontier energy, there may be a large array of supersymmetric particles
and,
if large extra dimensions exist,
mini black holes\,\cite{Godang}.
Like SPEAR, the resolution of a muon
collider is unaffected by beamstrahlung.
Muon ionization cooling is the key to this machine and a vigorous R\&D program is 
underway\,\cite{Torun,Palmer}.  Given a large initial emittance,
focusing magnets and RF cavities must be in close
proximity. Magnetic fields perpendicular to RF cavity surfaces enhance breakdown\,\cite{Norem}.
Possible cures include lattices with magnetic fields parallel to RF cavity surfaces
to bend electrons back into the cavity surface before they can accelerate,
high pressure hydrogen gas in RF cavities to slow electrons\,\cite{Yonehara}, grooved RF cavity walls
to trap electrons\,\cite{Wang}, or high melting point, low density materials such as beryllium 
to allow
sparks to spread their energy without melting RF cavity walls. 6D cooling guggenheims\,\cite{Klier} 
and rings\,\cite{RFOFO} show promise.
Final muon cooling requires short focal length lattices.  High T$_{\rm{c}}$ superconductors at 4K can carry large currents in  the 35 to 50\,T range\,\cite{Kahn}.   Parametric resonances\,\cite{Bogacz} 
and inverse cyclotrons\,\cite{Cyclotron} are also being explored for cooling.

\section{30 TO 400\,G{\lowercase{e}}V, 400\,H{\lowercase{z}} RING}
Historically synchrotrons have provided economical acceleration.  Here we outline 
a relatively fast 400\,Hz synchrotron\,\cite{Ramp}  for muons, which live for  2.2 $\mu$S.
A 200 $\mu$S (1250\,Hz) pulsed wiggler has been built with vanadium permen\-dur which is similar to the magnets required here. It achieved 2.1\,T in a 4\,mm gap with a 10\,cm wavelength\,\cite{Gallardo}.
In eq.\,1, the dipole vertical aperture, $h$, is calculated using an  emittance of 
$\epsilon_y\!=\!$ 25 $\pi$ mm-mrad\,\cite{Palmer}
and the Fermilab main ring FODO lattice parameter, $\beta_y\!=\!$ 99\,m.  Acceleration to 30 GeV might use dogbone recirculating LINACs and Fixed Field, Alternating Gradient (FFAG) rings\,\cite{Berg}. 

\vspace{-3mm}
\begin{equation}\label{eq:emit}
h = 6\sigma = 
6\sqrt{\frac{\epsilon_y \, \beta_y}  {6\pi \beta \gamma}} = 
6\sqrt{\frac{25\mu{\rm{m}} \ 99{\rm{m}}} {6\pi (1) (284)}} 
= 6\hbox{mm}
\end{equation}

Eqs.\,2 and 3 are now used to calculate the dipole voltages and amperages in Table 1.
$N$ is the number of turns in a coil.
A simple LC circuit with an IGBT or SCR  switch is used.
The voltages are reasonable because the magnetic field volume is small and little energy
is stored in the grain oriented 3\% silicon steel (Table 2). Achieving good field quality in small aperture
magnets needs to be explored.

\vspace{-3mm}
\begin{equation}\label{eq:joules}
W\!=\!\int\!\!\frac{B^2}{2\mu_0}\,dh\,{dw\,d\ell} =\! {\frac{LI^2}{2}} =\! {\frac{CV^2}{2}}, \   \
f =\!\frac{1}{2\pi\sqrt{LC}}
\end{equation}

\vspace{-3mm}
\begin{equation}\label{eg:amps}
I = B\,h / \mu_0N, \quad
V = 2 \pi \,B\, f\, N\, w\, \ell
\end{equation}

\begin{table}[hb!]
\begin{center}
\vspace{-5mm}
\caption{Fast ramping dipole parameters.}
\vspace{1mm}
\renewcommand{\arraystretch}{1.08}
\tabcolsep= 1.8mm
\begin{tabular}{lcccc} \hline
Injection energy & GeV   & 30  & 400  & 400  \\
Extraction energy &GeV & 400  & 750 & 750 \\    
Dipoles\,/\,half cell    &        &    4                        &  2                       &  1               \\            
Dipole  length, $\ell$ & m            &      6.3                   & 3.75                  &  7.5              \\ 
Bore height, $h$ & mm             &       6                      &     5                   &    5               \\
Bore width, $w$ & mm               &       30                    &   50                   &   50             \\
Initial magnetic field, $B$ &T                 &    0.14                    &  -1.8                 &   -1.8           \\
Final magnetic field, $B$ & T                  &     1.8                      &  1.8                  &   1.8            \\
Orbits                              &      &    28                       &   44                  &    44           \\
Acceleration period & ms  &    0.59                    &  0.92                &   0.92         \\
Frequency, $f$ & Hz                   &    400                     &   550                &   550          \\
Coil turns, $N$           &                   &     4                         &    4                         &    2              \\
Coil resistance, $R$ & $\mu\Omega$ & 4500            &     2700              &  1350      \\
Current, $I$ & A                    &     2200                  &    1800             &   3600       \\
Magnet energy, $W$ & J               &     1500                   &   1200              &    2400       \\
Magnet inductance, $L$ & $\mu$H  &   630                 &    760                      &   380                 \\
Capacitance, $C$ &  $\mu$F             &     250                 &    110                      &   220                \\
Voltage, $V$ &   V                                   &   3400                 &      4700                   & 4700      \\
Power Consumption             & kW      &           0.6                    &    1.4                        &  2.8         \\
\hline           
\end{tabular}
\label{dipoles}
\end{center}
\end{table}

\smallskip
Using eq.\,4, the skin depth, $\delta$, of steel with $\mu\!=\!3000\mu_0$ at 400\,Hz is 0.3\,mm.
From eq.\,5, only 2\% of $t\!=\!$ 0.28\,mm thick steel is lost due to shielding by eddy currents\,\cite{Scott}.
The skin depth for 18\,n$\Omega$-m copper at 400\,Hz is 3.4\,mm.
\vspace{-4mm}

\begin{equation}\label{eq:skin}
\delta = \sqrt{\rho  /  (\pi \, f \,\mu)} =
\sqrt{470\!\times\!{10^{-9}}  /  (\pi \, 400 \, (3000\,\mu_0))}
\end{equation}
\vspace{-10mm}

\vspace{-1mm}
\begin{equation}
{\hbox{L/L}}_0 = \frac{(\delta/t) \, (\sinh(t/\delta) + \sin(t/\delta))}
{\cosh(t/\delta) + \cos(t/\delta)} = 0.98
\end{equation}

Now we estimate the power consumption of the magnets.
Laminations are  laid out to minimize core losses (Fig. 1).
Eq.\,6\,\cite{McLyman} gives a value of 23\,W/kg for the steel.
 An average magnetic field of 1.6\,T is used.  Both eddy currents and hysteresis  losses,
$\int{\bf{H}}{\cdot}d\,{\bf{B}}$, which scale
with the coercive force,
H$_c$, given in Table 2, are included.  Eddy currents alone\,\cite{Sasaki}  give 15\,W/kg in eq.\,7.
The total core loss for a one ton dipole is 23\,kW. $I^2R$ losses for four turns of $1\!\times\!2$\,cm 
copper  (4500 $\mu\Omega$) carrying 2200\,A of sinusoidal current are 11\,kW.  
Using eq.\,7 with an 0.1\,T
field and coils made of 2\,mm transposed strands, the eddy current losses in the copper are
6\,kW.  So multiplying 40\,kW per dipole times 800 dipoles and adding 6\% for quadrupoles, one gets
34000\,kW.  But the magnets are only on for half of a 400\,Hz cycle, 13 times per second,
for a duty cycle of  1.6\%  and a total power consumption of 540\,kW. A choke and
diode are used to do a leisurely  reset of the polypropylene capacitor bank polarity for each new cycle.      

\vspace{-5.5mm}
\begin{equation}\label{eq:core}
\hbox{Core Loss}\,=\, 4.38\!\times\!10^{-4}
f^{1.67}
B^{1.87} \,=\,23\,\hbox{W/kg}
\end{equation}

\vspace{-4.5mm}
\begin{equation}
P = \hbox{[Volume]}{{(2\pi\,f\,B\,w)^2}\over{24\rho}}
= 15\,\hbox{W/kg}
\end{equation}

\begin{table}[hbt]
\begin{center}
\vspace{-1mm}
\caption{
Resistivity ($\rho$), coercivity (H$_{\rm{c}}$), and  permeability ($\mu/\mu_0$) of steels.
Higher resistivity lowers eddy current losses. Low coercivity minimizes hysteresis losses.
Grain oriented 3\% silicon steel has a far
higher permeability parallel ($\parallel$) to
than perpendicular ($\perp$) to
its rolling direction \cite{Bozorth}
and permits minimal
energy ($B^2/2\mu$) storage in the yoke, as compared to low carbon steel\,\cite{LHC} at 1.8\,T. }
\vspace{2.0mm}
\renewcommand{\arraystretch}{1.08}
\tabcolsep= 0.8mm
\begin{tabular}{lccrrr} \hline
Steel                           & $\rho({\rm{n}}\Omega$-m) & H$_{\rm{c}}$(A/m)   &  1.0\,T & 1.5\,T & 1.8\,T \\ \hline
.0025\% Carbon   & 100  &  80  &    4400 &   1700 &  240   \\
Oriented ($\parallel$) Si & 470&  8 & 40000 &  30000 & 3000   \\
Oriented ($\perp$)  Si  &  470  & & 4000 & 1000   &       \\ \hline
\end{tabular}
\label{steel}
\end{center}
\vspace{-12mm}
\end{table}

\begin{figure}[t]
\centering
\includegraphics*[width=65mm]{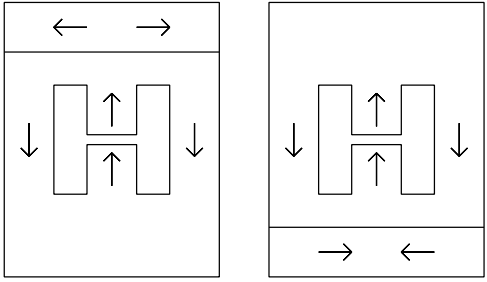}
\vspace{-3mm}
\caption{Alternating dipole laminations of grain oriented silicon steel.
The arrows show the {\bf B} field direction and the grain direction.
The layout resembles  an ``EI" transformer.}
\vspace{-3mm}
\end{figure}

\begin{figure}[b]
{\vspace{-1.8cm}\hspace{-1.7cm}\rotatebox{90}
{\includegraphics*[width=85mm]{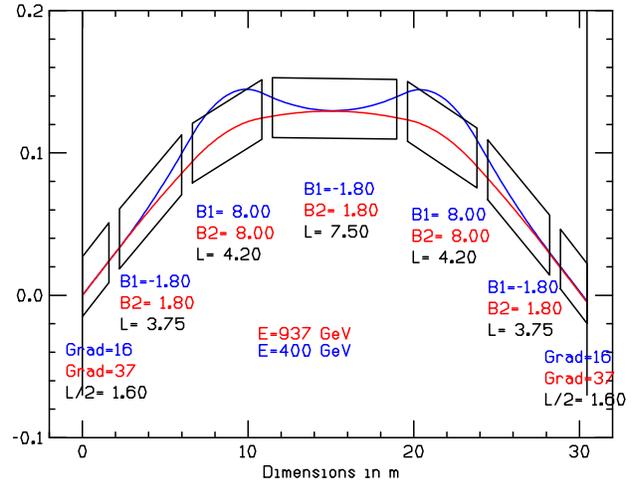}}}
\vspace{-18mm}
\caption{The 30.45\,m  long FODO lattice half cell consists of half of two 
3.2\,m ramping quadrupoles, two fixed, 4.2\,m 8\,T superconducting dipoles, and two 3.75\,m plus one
 7.5\,m ramping dipoles.  The ramping dipoles, which go from  -1.8\,T to 1.8\,T at 550 Hz,
 oppose the superconducting dipoles at \textcolor{blue}{400 GeV} and act in unison at 
 \textcolor{red}{750 GeV} or perhaps a bit higher.}
\label{half-cell}
\end{figure}

\section{400 TO 750\,G{\lowercase{e}}V, 550\,H{\lowercase{z}}  HYBRID RING}

The 400\,GeV Fermilab main ring FODO lattice is slightly modified to reach 750\,GeV.   The 30 T/m quadrupoles are lengthened from  1.7\,m to 3.2\,m and run at 150\,Hz.  The four ramping dipoles per half cell are replaced by five dipoles, two fixed 8\,T superconducting dipoles in between three dipoles ramping from -1.8\,T to 1.8\,T at 550\,Hz, as shown in Fig.\,2.  The ramping dipoles oppose the superconducting dipoles at injection and work in  unison at extraction.  Ramping dipole parameters are given in the last two columns of Table 1.

Now we estimate the power consumption of the magnets.
Using an  average magnetic field of 1.6\,T  and a frequency of 550\,Hz, 
Eq.\,6\,\cite{McLyman} gives a value of 40\,W/kg for the 0.28\,mm grain oriented 3\% silicon steel.
The total core loss for a 2400\,kg, 7.5\,m long dipole is 96\,kW. $I^2R$ losses for two turns of 
$2\!\times\!2$\,cm 
copper  (1350 $\mu\Omega$) carrying 3600\,A of sinusoidal current is 9\,kW for the 7.5\,m long dipole.  Using eq. 7 with an 0.1\,T
field and coils made of 2\,mm transposed strands, the eddy current losses in the copper are
13\,kW.  So multiplying 118\,kW per 7.5\,m dipole times 200 7.5\,m and 400  3.75\,m  dipoles and adding 6\% for quadrupoles, one gets
50000\,kW.  But the magnets are only on for a 550\,Hz cycle, 13 times per second,
for a duty cycle of  2.4\%  and a total power consumption of 1200\,kW.

\section{1.3\,GH{\lowercase{z}}, 10\,MW KLYSTRONS}

Acceleration from 30 to 400\,GeV uses 14 GV  of 1.3\,GHz superconducting RF
in 42 locations evenly spaced around the ring. 
The acceleration occupies 0.59\,ms and  28 orbits.  Muon survival is 80\%.
Forty-two 10 MW klystrons allow the energy extracted  by a pair of $2 \times 10^{12}$ muon bunches to be replaced.  One RF coupler for every three cells is required.

Acceleration from 400 to 750\,GeV uses 8\,GV of 1.3\,GHz superconducting RF in 12 locations evenly spaced around the ring. 
The acceleration occupies 0.92\,ms and 44 orbits.  Muon survival is 92\%.
Twenty-four 10 MW klystrons allow the energy taken by a pair of $2 \times 10^{12}$ muon bunches to 
be replaced.  One RF coupler for every three cells is required.

Running at 13\,Hz, the cryogenics and klystron modulators require 4 and 22\,MW  of AC wall power, respectively.

A bunch with $2 \times 10^{12}$ muons extracts 8\% of the energy from an RF cavity leading to
head/tail, wakefield\,\cite{Neuffer}, and HOM\,\cite{Padamsee} issues.  
One cell stores 13 joules at 31.5 MV/m.
However, as shown in eqs.\,8 and 9,
there are synchrotron oscillations\,\cite{Courant}  to aid longitudinal dynamics. $h$ is the harmonic number (number of 0.23\,m RF wavelengths around the ring).  The transition $\gamma$
is 18 for the main ring, which gives a momentum compaction, $\eta$, 
of $\frac{1}{\rule[0pt]{0pt}{6pt}18^2}$. 

\vspace{-2mm}
\begin{equation}\label{eq:transition}
d\tau/\tau =  (1/\gamma_t^2 - 1/\gamma^2) (dp/p) = \eta (dp/p)
\end{equation}

\vspace{-6mm}
\begin{equation}\label{eq:oscillation}
\nu_s\! =\! 
\sqrt{\frac{h\eta\rule[-4pt]{0pt}{10pt}(\rm GV)\cos{\phi_s}}{-2\pi\beta^2E_s}}\!
=\!
\sqrt{\frac{27200\rule[-4pt]{0pt}{10pt} (\frac{\rule[-2pt]{0pt}{5pt}1}{\rule[-2pt]{0pt}{8pt}18^2})14(.1)}
{2\pi(1^2)(30)}}\! =\! .8
\end{equation}

Muon bunches must stay in phase with the RF.
The muon speed increase from $\beta\! =\,$0.99999380 to $\beta\! =\,$0.99999996 
in the first ring can be corrected by increasing the orbital radius by 6\,mm
during acceleration.  The one in  40000 path length decrease in the second ring
can be corrected by increasing the orbital radius by 25\,mm during acceleration. 

A longitudinal emittance of 0.072\,$\pi$\,m-rad\,\cite{Palmer} leads to an 0.01 m long
muon bunch injected at 30\,GeV/c with a 2.5\% momentum spread, 
i.e. $0.072\!=\!(0.025(30)/m_{\mu^{\pm}})(0.01)$,
where   $m_{\mu^{\pm}}\! =\! 0.106\,\hbox{GeV/c}^2$.
Better might be 0.005\,m and 2\%\,\cite{Neuffer}.  
The RF wavelength is 0.23\,m. A muon on crest gets 4\% more acceleration 
than one 0.01\,m (15$^0$) off crest.

Using eq.\,10, and integrating over the acceleration cycle from 30 to 750 GeV with
$4 \times 10^{12}$ muons at 13\,Hz, and neglecting downtime and straight sections in the ring, a person would receive
a dose of a millirem at 2700\,m from decays into neutrinos, if they stood in the beam constantly.
This is 1\% of the federal limit, 10\% of the Fermilab offsite limit, and equivalent to eating two bananas a week.
Note that the Fox River is 5000\,m away from and 4\,m below the Tevatron, so neutrinos at  2700\,m are still underground.

\vspace{-4mm}
\begin{equation}
\hbox{distance(meters)} = 5 \times 10^{-7} \sqrt{\mu/\hbox{year}}\, \, \hbox{E\,(TeV)}^{1.5}
\end{equation}

Many thanks to A. Garren, 
D. Trbojevic, K. Bourkland, D. Wolff, R. Rimmer, D. Li, H. Padamsee, M. Syphers, S.~Kahn,
and D. Neuffer for sage advice.

  
\end{document}